\documentclass[a4paper,11pt]{article}
\usepackage{jheppub} 
\usepackage{lineno}

\arxivnumber{2409.05588} 

\title{CP violation in two-meson Tau decays induced by heavy new physics}

\author[1]{Daniel A. L\'opez Aguilar,}
\author[2]{Javier Rend\'on,}
\author[1,3]{and Pablo Roig}
\affiliation[1]{Departamento de Física, Centro de Investigación y de Estudios Avanzados del Instituto Politécnico Nacional, Apdo. Postal 14-740, 07000 Ciudad de México, México.}
\affiliation[2]{Instituto de Ciencias Nucleares, Universidad Nacional
Autónoma de México, A.P. 70-543, Ciudad de México, México.}
\affiliation[3]{IFIC, Universitat de Val\`encia – CSIC, Catedr\'atico Jos\'e Beltr\'an 2, E-46980 Paterna, Spain.}

\emailAdd{Daniel.Lopez.Aguilar@cinvestav.mx} \emailAdd{jesus.rendon@correo.nucleares.unam.mx} 
\emailAdd{Pablo.Roig@cinvestav.mx} 
\emailAdd{paroig@ific.uv.es}

\abstract{We apply the effective field theory formalism that was used to study CP violation induced by heavy new physics in the $\tau\to K_S \pi\nu_\tau$ decays to the other two-meson tau decay channels. We focus on the rate and the forward-backward asymmetries, that are predicted using current bounds on the complex Wilson coefficients of the effective Lagrangian. We discuss our outcomes for the modes with $(K/\pi)\pi^0$ and $K K_S$, that can be studied at Belle-II and a super-tau-charm facility. Our main finding is that current and forthcoming experiments would be sensitive to the maximum allowed CP rate asymmetry in the $K^\pm K_S$ modes if a precision of $5\%$ is reached on this observable, that can check as well the BaBar anomaly in $K_S\pi$. For the $\pi^\pm\pi^0$ channels, new physics would be difficult to probe at present. Disentangling new sources of CP violation would be most challenging in $K^\pm\pi^0$ and the other modes.}

\begin{document}
\maketitle
\flushbottom

\section{Introduction}\label{sec:Intro}
\hspace{0.5cm}Violation of the invariance under the combined symmetry operations of charge conjugation (C) and parity (P), CP, is a required ingredient -within any Lorentz invariant local quantum field theory with a Hermitian Hamiltonian- to understand the baryon asymmetry of the Universe, as pointed out by Zakharov in his seminal paper \cite{Sakharov:1967dj}. It also needs baryon number violation (which could be triggered violating initially the lepton number~\cite{Fukugita:1986hr}), non-conservation of C, and interactions out of thermal equilibrium. In principle all these conditions can be met within the Standard Model (SM)~\cite{Glashow:1961tr, Weinberg:1967tq, Salam:1968rm}, though it fails to explain the current matter-antimatter asymmetry~\cite{ParticleDataGroup:2024cfk}. This conundrum enhances the case for searches of additional sources of CP violation, that can eventually solve this puzzle.\\

In the quark sector, CP violation in the SM arises only from an irreducible phase in the Cabibbo-Kobayashi-Maskawa (CKM) mixing matrix \cite{Cabibbo:1963yz,Kobayashi:1973fv}, and has been measured in $K$, $B_{(s)}$ and $D$ mesons \cite{Christenson:1964fg,NA31:1988eyf,KTeV:1999kad,NA48:1999szy,BaBar:2001pki,Belle:2001zzw,BaBar:2004gyj,Belle:2004nch,LHCb:2012xkq,LHCb:2013syl,LHCb:2019hro}. To date, no data from hadron decays or meson mixing seems to require CP violation beyond the SM~\cite{ParticleDataGroup:2024cfk}. Conversely, CP violation in the lepton sector has not been discovered yet, although there are hints for it both in neutrino oscillation measurements by Nova \cite{NOvA:2021nfi} and T2K \cite{T2K:2023smv}, and in global fits~\cite{ParticleDataGroup:2024cfk}. If confirmed, it would remain to be seen if it can be explained by the single CP violating phase that appears in the lepton analogue of the CKM matrix, the Pontecorvo-Maki-Nakagawa-Sakata matrix \cite{Maki:1962mu,Pontecorvo:1967fh,Katayama:1962mx}, or new physics is called for.\\

Hadron tau decays are not only a privileged arena for learning about non-perturbative QCD in a clean environment, but also a very useful tool in the quest for new physics~\cite{Pich:2013lsa, Davier:2005xq}. Regarding CP violation, this was emphasized for the first generation B-factories, BaBar and Belle, and forthcoming facilities~\cite{Tsai:1996ps, Kuhn:1996dv, Bigi:2012km, Bigi:2012kz, Kiers:2012fy}. CP non-conservation in semileptonic tau decays was boosted by the anomaly reported by BaBar \cite{BaBar:2011pij} on the difference between the decay rates of a $\tau^+$ and a $\tau^-$ into final states with a correspondingly charged pion and a $K_S$ (inclusive in possible neutral pions), normalized by their sum. While this CP violating observable should be $(0.36\pm0.01)\%$ in the Standard Model --due to neutral kaon mixing and accounting for the precise BaBar experimental setting--~\cite{Bigi:2005ts,Calderon:2007rg,Grossman:2011zk}, the measurement came out with opposite sign, $-(0.36 \pm 0.23 \pm 0.11)\%$ (with statistical and systematic errors in turn), being $2.8$ standard deviations away from the theory prediction. Although there were claims that this effect could be due to heavy new physics (particularly of tensor type \cite{Devi:2013gya,Dhargyal:2016kwp,Dighe:2019odu}), the key equality of the tensor and vector form factors phases in the elastic region allowed to disfavor this explanation unless an unnatural fine-tuning was involved \cite{Cirigliano:2017tqn} (see also refs.~\cite{Rendon:2019awg,Chen:2019vbr,Chen:2020uxi,Sang:2020ksa,Chen:2021udz}).~\footnote{No CP violation was found in the binned analysis of these decays by Belle~\cite{Belle:2011sna}. However, its precision was not enough to clarify the BaBar anomaly (neither in the pioneering CLEO studies~\cite{CLEO:1998lwx,CLEO:2001lhp}).}\\

The aim of this paper is to extend the work done in the previous references to the other two-meson tau decays of interest ($\pi\pi$, $K\bar{K}$ and $K\pi^0$ channels).~\footnote{The $\pi\eta^{(\prime)}$ channels have not been measured yet, and the hadronic uncertainty in them is big~\cite{Descotes-Genon:2014tla, Escribano:2016ntp}, so any current prediction would be subject to large uncertainties. CP conserving new physics imprints in these decays were the subject of ref.~\cite{Garces:2017jpz}. We will comment briefly on the $K(\eta/\eta')$ channels, also subject to large uncertainties, at the end of section \ref{sec_Pheno}.}~\footnote{Model-dependent studies have been carried out in e.~g. refs.~\cite{Delepine:2006fv,Delepine:2007qg,LopezCastro:2009da,Kimura:2012bwp,Gao:2012su,Delepine:2018amd,Faisel:2023ldd}.}~\footnote{We will not address in this work the case of polarized taus, which would allow further tests of $CP$ and time-reversal symmetry violations, along the lines of ref.~\cite{Chen:2022nxm}.} In the same way that heavy new physics contributions to the CP violating $\tau\to K_S\pi\nu_\tau$ observables is very much restricted, model-independently, upon translating the constraints on the SM effective-field theory (EFT) \cite{Buchmuller:1985jz, Grzadkowski:2010es} operators to the low-energy EFT (LEFT) ones, it will happen similarly in the other two-meson tau decay modes. Our purpose is to compute the maximum amount of CP violation that can be due to heavy new physics in these processes. In this way, any observation beyond our predicted upper limits~\footnote{Within the Standard Model CP violation is negligibly small in all tau decays without a neutral Kaon, see for instance ref.~\cite{Delepine:2005tw}.} will either be due to light new physics or signal underestimated uncertainties, which would deserve a close scrutiny. The discussed physics shall be of interest for Belle-II \cite{Belle-II:2018jsg} and future super-tau-charm factory \cite{Achasov:2023gey, Achasov:2024eua,CPVatSTCF} searches.\\

Our paper is structured as follows. In section \ref{sec:LEFTCPVObservables} we give the LEFT Lagrangian for the $\tau^-\to \bar{u} (d/s) \nu_\tau$ decays and quote the relevant matrix element, where hadronization is encoded in a set of scalar, vector and tensor form factors, which have been worked out in previous articles, as sketched in section \ref{sec:FFinput}. In section \ref{sec:LEFTCPVObservables} we also give the doubly differential decay rate and the expressions obtained by integrating upon the relevant angle and the invariant mass of the meson system, in turn. The latter expression allows us to find straightforwardly the CP asymmetry rate, which is one of the central observables considered in our study. We also recall how to obtain the other main observable, the forward-backward asymmetry, from the doubly differential distribution of the decay width, and include other useful results. In section \ref{sec:BoundsonEpsilons} we review the known information on the complex Wilson coefficients of the effective Lagrangian, which will determine the possible NP signals, that we analyze in section \ref{sec_Pheno}. Finally, we state our conclusions in section \ref{sec_Concl}.

\section{Low-energy Effective Field Theory Computation}\label{sec:LEFTCPVObservables}
\hspace{0.5cm} The most general Fermi-like effective Lagrangian for the $\tau^-\to\bar{u}D\nu_\tau\, (D=d,s)$ decays is given by \cite{Cirigliano:2009wk,Garces:2017jpz,Cirigliano:2017tqn}~\footnote{In our conventions, $\sigma^{\mu\nu}=\frac{i}{2}[\gamma^\mu,\gamma^\nu]$ and $\epsilon^{0123}=+1$.}
\begin{eqnarray}\label{LEFTLagrangian}
\mathcal{L}_{
EFT}=-\frac{G_F^0V_{uD}}{\sqrt{2}}\left(1+\epsilon_L^{D}+\epsilon_R^{D}\right)\Bigg\lbrace\bar{\tau}\gamma_\mu(1-\gamma_5)\nu_\tau\cdot\bar{u}\left[\gamma^\mu-\left(1-2\hat{\epsilon}_R^{D}\right)\gamma^\mu\gamma_5\right]D\nonumber\\
+\bar{\tau}(1-\gamma_5)\nu_\tau\cdot\bar{u}\left[\hat{\epsilon}_S^{D}-\hat{\epsilon}_P^{D}\gamma_5\right]D+2\hat{\epsilon}_T^{D}\bar{\tau}\sigma_{\mu\nu}(1-\gamma_5)\nu_\tau\cdot\bar{u}\sigma^{\mu\nu}D\Bigg\rbrace+\mathrm{h.c.},
\end{eqnarray}
where $G_F^0$ is the Fermi constant in the SM, $V_{uD}$ is the corresponding CKM matrix element, and the $\epsilon_i\,(i=S,P,V,A,T)$ are the complex Wilson coefficients~\footnote{An overall phase is non-physical.} parameterizing the interactions mediated by (integrated-out) heavy new physics, in such a way that setting $\epsilon_i=0$ we recover the SM contributions. We will use the $\epsilon_i$ at the renormalization scale $\mu=2$ GeV and work in the $\overline{MS}$ scheme (so we will proceed in this way also for the quark masses). We will employ $G_F=G_F^0(1+\epsilon_L+\epsilon_R)$ to denote the experimentally determined value of the Fermi constant, which will motivate our definition of the $\hat{\epsilon}_i=\epsilon_i/(1+\epsilon_L+\epsilon_R)$. Our results will be valid at linear order in the small ($\hat{\epsilon}_i\sim \epsilon_i$) $\epsilon_i$.\\

The corresponding decay amplitude for the $\tau^-\to K^- \pi^0 \nu_\tau$ decays is (equations of motion were applied)
\begin{equation}\label{matrixelement}
\mathcal{M}(\tau^-\to K^- \pi^0 \nu_\tau)=\frac{G_FV_{us}}{2}\left[L_\mu H^\mu+\hat{\epsilon}_S^*LH+2\hat{\epsilon}_T^*L_{\mu\nu}H^{\mu\nu}\right]\,,
\end{equation}
where the lepton and hadron currents are defined as
\begin{eqnarray}
   L&=&\bar{u}(p_{\nu_\tau})(1+\gamma_5)u(p_\tau)\,,\\
   L_\mu&=&\bar{u}(p_{\nu_\tau})\gamma_\mu(1-\gamma_5)u(p_\tau)\,,\\
   L_{\mu\nu}&=&\bar{u}(p_{\nu_\tau})\sigma_{\mu\nu}(1+\gamma_5)u(p_\tau)\,,
\end{eqnarray}
and
\begin{eqnarray}
H&=&\left\langle\pi^0(p_\pi)K^-(p_K)|\bar{s}d|0\right\rangle=\frac{\Delta_{K\pi}}{m_s-m_u}F_0(s)\,,\\
H^\mu&=&\left\langle\pi^0(p_\pi)K^-(p_K)|\bar{s}\gamma^\mu d|0\right\rangle=\left[(p_\pi-p_K)^\mu+\frac{\Delta_{K\pi}}{s}q^\mu\right]F_+(s)-\frac{\Delta_{K\pi}}{s}q^\mu F_0(s)\,,\;\;\;\;\;\;\\
H^{\mu\nu}&=&\left\langle\pi^0(p_\pi)K^-(p_K)|\bar{s}\sigma^{\mu\nu} d|0\right\rangle=iF_T(s)(p_\pi^\mu p_K^\nu-p_K^\mu p_\pi^\nu)\,,
\end{eqnarray}
in which $q^\mu=(p_\pi+p_K)^\mu$, $s=q^2$, $\Delta_{K\pi}=m_K^2-m_\pi^2$, and $F_{0,+,T}(s)$ are, respectively, the scalar, vector and tensor form factors, encoding the corresponding hadronization of the quark currents between the QCD vacuum and the final-state mesons.\\

The other decay channels of interest are trivially obtained from the $K^-\pi^0$ one, replacing $p_K$ ($p_\pi$) by the charged (neutral) meson momentum and substituting the $1/2$ factor in eq.~(\ref{matrixelement}) by $1/\sqrt{2}$ for the $K^-K^0$ and $\bar{K}^0\pi^-$ modes and by $1$ in the $\pi^-\pi^0$ channel. The $K^-\eta^{(\prime)}$ channels are a bit more complicated. Relative to the $K^-\pi^0$ modes, one needs to substitute $1\to c^V_{K^-\eta^{(\prime)}}$ in the $F_+(s)$ term and $1\to-c^S_{K^-\eta^{(\prime)}}$ in the $F_0(s)$ term, where $c^V_{K^-\eta^{(\prime)}}=-\sqrt{\frac{3}{2}}$, $c^S_{K^-\eta}=-\frac{1}{\sqrt{6}}$ and $c^S_{K^-\eta^\prime}=\frac{2}{\sqrt{3}}$~\cite{Escribano:2013bca}. The scalar matrix element, $H$, demands to modify $1\to c^S_{K^-\eta^{(\prime)}}$, in order to obtain the $K^-\eta^{(\prime)}$ result from the $K^-\pi^0$ expression. For the tensor matrix element $H^{\mu\nu}$, the $K^-\eta^{(\prime)}$ result is found from the $K^-\pi^0$ one by changing $1\to c^V_{K^-\eta^{(\prime)}}$.

The observables of interest are quoted in the following. All of them check the results in ref.~\cite{Chen:2021udz} accounting for the relative factor of $1/2$ entering our results for the (differential) decay width:
\begin{eqnarray}
 &&\frac{\mathrm{d}^2\Gamma(\tau^-\to K^-\pi^0\nu_\tau)}{\mathrm{d}s\,\mathrm{d}\cos\alpha}=\frac{G_F^2|V_{us}|^2M_\tau^3S_{EW}}{1024\pi^3s^3}\left(1-\frac{s}{M_\tau^2}\right)^2\lambda^{1/2}(s,m_K^2,m_\pi^2)\nonumber\\
 &&\times\Bigg\lbrace\lambda(s,m_K^2,m_\pi^2)\left[\frac{s}{M_\tau^2}+\left(1-\frac{s}{M_\tau^2}\right)\cos^2\alpha\right]|F_+(s)|^2+\Delta_{K\pi}^2\Bigg|1+\frac{\hat{\epsilon}_S s}{M_\tau(m_s-m_u)}\Bigg|^2|F_0(s)|^2\nonumber\\
&&+4\lambda(s,m_K^2,m_\pi^2)\left[s|\hat{\epsilon}_T|^2\left(1-\left(1-\frac{s}{M_\tau^2}\right)\cos^2\alpha\right)|F_T(s)|^2-\frac{s}{M_\tau^2}\Re e[\hat{\epsilon}_TF_+(s)F^*_T(s)]\right]\nonumber\\
&&-2\Delta_{K\pi} \lambda^{1/2}(s,m_K^2,m_\pi^2)\Re e\left[\left(1+\frac{\hat{\epsilon_S s}}{M_\tau(m_s-m_u)}\right)F_+(s)F_0^*(s)\right]\cos\alpha\nonumber\\
&&+\frac{4s}{M_\tau}\Delta_{K\pi}\lambda^{1/2}(s,m_K^2,m_\pi^2)\Re e\left[\hat{\epsilon}^*_T\left(1+\frac{\hat{\epsilon}_S s}{M_\tau(m_s-m_u)}\right)F_T(s)F_0^*(s)\right]\cos\alpha\Bigg\rbrace\,, 
\end{eqnarray}
where $\lambda(a,b,c)=a^2+b^2+c^2-2(ab+ac+bc)$ and $S_{EW}\sim1.02$~\cite{Marciano:1988vm} encodes the universal short-distance electroweak radiative corrections (which only affect the SM contribution, but are written here as an overall factor for simplicity, since the induced error of this approximation is subleading).~\footnote{We are not considering observables inclusive in photon emission, so that the process-dependent long-distance electromagnetic correction~\cite{Cirigliano:2001er, Cirigliano:2002pv, Miranda:2020wdg} is not applied.} Integrating the angular dependence,~\footnote{We recall that $\alpha$ is the angle between one of the mesons momentum and the tau in the hadronic rest frame \cite{Kuhn:1992nz, Belle-II:2018jsg}. } we get
\begin{eqnarray}
&&\frac{\mathrm{d}\Gamma(\tau^-\to K^-\pi^0\nu_\tau)}{\mathrm{d}s}=\frac{G_F^2|V_{us}|^2M_\tau^3S_{EW}}{1024\pi^3s^3}\left(1-\frac{s}{M_\tau^2}\right)^2\lambda^{1/2}(s,m_K^2,m_\pi^2)\nonumber\\
&&\times\Bigg\lbrace \frac{2}{3}\lambda(s,m_K^2,m_\pi^2)\left(1+\frac{2s}{M_\tau^2}\right)|F_+(s)|^2+2\Delta_{K\pi}^2\Bigg|1+\frac{\hat{\epsilon}_S s}{M_\tau(m_s-m_u)}\Bigg|^2|F_0(s)|^2\nonumber\\
&&+\frac{8}{3}\lambda(s,m_K^2,m_\pi^2)\left(s|\hat{\epsilon}_T|^2\left(2+\frac{s}{M_\tau^2}\right)|F_T(s)|^2-\frac{3s}{M_\tau}\Re e[\hat{\epsilon}_TF_+(s)F^*_T(s)]\right)\Bigg\rbrace\,.\;\;\;\;\;\;\;\;\;
\end{eqnarray}
Integrating upon $s$ we obtain the partial decay width for the considered channel, that is also needed in the following.\\

Of particular interest will be the rate CP asymmetry, given by
\begin{eqnarray}\label{eq_ACPrateNP}
A^{\mathrm{rate}}_{CP}(\tau^\pm\to K^\pm\pi^0\nu_\tau)&=&\frac{\Gamma(\tau^+\to K^+\pi^0\bar{\nu}_\tau)-\Gamma(\tau^-\to K^-\pi^0\nu_\tau)}{\Gamma(\tau^+\to K^+\pi^0\bar{\nu}_\tau)+\Gamma(\tau^-\to K^-\pi^0\nu_\tau)}\nonumber\\
&=&\frac{\Im m[\hat{\epsilon}_T]G_F^2|V_{us}|^2S_{EW}}{128\pi^3M_\tau^2\Gamma(\tau\to K\pi^0\nu_\tau)}\int_{(m_K+m_\pi)^2}^{M_\tau^2}\mathrm{d}s\left(1-\frac{M_\tau^2}{s}\right)^2\lambda^{3/2}(s,m_K^2,m_\pi^2)\nonumber\\
&& \times |F_T(s)||F_+(s)|\sin[\delta_T(s)-\delta_+(s)]\,,
\end{eqnarray}
where $\delta_{+,T}(s)$ denote the phases of the vector and tensor form factors, in turn. The expression in the second and third lines of eq.~(\ref{eq_ACPrateNP}) corresponds to the new physics (NP) contribution to $A_{CP}^{rate}$, which coincides with $A_{CP}^{rate}$ for those modes without neutral Kaons. In the $\tau^\pm\to K^\pm K_S\nu_\tau$ case, this observable is dominated by the SM contribution, according to~\cite{Devi:2013gya}
\begin{equation}\label{eq_FullACPrate}
A_{CP}^{rate}\,=\,\frac{A_{CP}^{rate}|_{SM}+A_{CP}^{rate}|_{NP}}{1+A_{CP}^{rate}|_{SM}\times A_{CP}^{rate}|_{NP}}\,.
\end{equation}
\\
\indent Another useful observable is the forward-backward asymmetry,
\begin{equation}
A_{FB}^{\tau^-\to K^-\pi^0\nu_\tau}(s)=\frac{\int_0^1\frac{\mathrm{d}^2\Gamma(\tau^-\to K^-\pi^0\nu_\tau)}{\mathrm{d}s\,\mathrm{d}\cos\alpha}\mathrm{d}\cos\alpha-\int_{-1}^0\frac{\mathrm{d}^2\Gamma(\tau^-\to K^-\pi^0\nu_\tau)}{\mathrm{d}s\,\mathrm{d}\cos\alpha}\mathrm{d}\cos\alpha}{\int_0^1\frac{\mathrm{d}^2\Gamma(\tau^-\to K^-\pi^0\nu_\tau)}{\mathrm{d}s\,\mathrm{d}\cos\alpha}\mathrm{d}\cos\alpha+\int_{-1}^0\frac{\mathrm{d}^2\Gamma(\tau^-\to K^-\pi^0\nu_\tau)}{\mathrm{d}s\,\mathrm{d}\cos\alpha}\mathrm{d}\cos\alpha}\,,
\end{equation}
that fulfills $A_{FB}(s)=3/2<\cos\alpha>(s)$, with
\begin{equation}
<\cos\alpha>(s)=\frac{\int_{-1}^1\cos\alpha\left(\frac{\mathrm{d}^2\Gamma(\tau^-\to K^-\pi^0\nu_\tau)}{\mathrm{d}s\,\mathrm{d}\cos\alpha}\mathrm{d}\cos\alpha\right)}{\int_{-1}^1\left(\frac{\mathrm{d}^2\Gamma(\tau^-\to K^-\pi^0\nu_\tau)}{\mathrm{d}s\,\mathrm{d}\cos\alpha}\mathrm{d}\cos\alpha\right)}=\frac{N(s)}{D(s)},
\end{equation}
where
\begin{eqnarray}\label{eq_N(s)}
    N(s)=&-&\frac{4}{3}\Delta_{K\pi}\lambda^{1/2}(s,m_K^2,m_\pi^2)\Re e\left[\left(1+\frac{\hat{\epsilon}_S s}{M_\tau(m_s-m_u)}\right)F_+(s)F_0^*(s)\right]\\
    &+&\frac{8s}{3M_\tau}\Delta_{K\pi}\lambda^{1/2}(s,m_K^2,m_\pi^2)\Re e\left[\hat{\epsilon}^*_T\left(1+\frac{\hat{\epsilon}_S s}{M_\tau(m_s-m_u)}\right)F_T(s)F_0^*(s)\right]\,,\;\;\;\;\;\;\;\nonumber\\
    D(s)&=&\frac{2}{3}\lambda(s,m_K^2,m_\pi^2)\left(1+\frac{2s}{M_\tau^2}\right)|F_+(s)|^2+2\Delta_{K\pi}^2\Bigg|1+\frac{\hat{\epsilon}_S s}{M_\tau(m_s-m_u)}\Bigg|^2|F_0(s)|^2\nonumber\\
    &+&\frac{8}{3}\lambda(s,m_K^2,m_\pi^2)\left[s|\hat{\epsilon}_T|^2\left(2+\frac{s}{M_\tau^2}\right)|F_T(s)|^2-\frac{3s}{M_\tau}\Re e[\hat{\epsilon}_TF_+(s)F_T^*(s)]\right]\,.\label{eq_D(s)}
\end{eqnarray}
From eqs.~(\ref{eq_N(s)}) and (\ref{eq_D(s)}) it is clear that $A_{FB}$ vanishes in the flavor symmetry limit and that it is not directly sensitive to CP violation, since the leading dependence on the new physics enters proportional to $\Re e[\hat{\epsilon}_{S,T}^{(\star)}F_{+,T}(s)F_0^*(s)]$. Then, measuring $A_{FB}(s)$ -or any observable directly linked to it- different from the SM prediction will require additional work to disentangle the effects due to CP violation.\\

For completeness, we note that these observables can be made bin-dependent straightforwardly, following ref.~\cite{Chen:2021udz}, namely
\begin{equation}
<\cos\alpha>_i^{\tau^-}=\frac{\int_{s1,i}^{s2,i}\int_{-1}^{+1}\cos\alpha\left(\frac{\mathrm{d}^2\Gamma(\tau^-\to K^-\pi^0\nu_\tau)}{\mathrm{d}s\,\mathrm{d}\cos\alpha}\mathrm{d}s\,\mathrm{d}\cos\alpha\right)}{\int_{s1,i}^{s2,i}\int_{-1}^{+1}\left(\frac{\mathrm{d}^2\Gamma(\tau^-\to K^-\pi^0\nu_\tau)}{\mathrm{d}s\,\mathrm{d}\cos\alpha}\mathrm{d}s\,\mathrm{d}\cos\alpha\right)}\,,
\end{equation}
and similarly for $\tau^-\to\tau^+$, doing $(\tau/K)^-\to(\tau,K)^+$, $\nu_\tau\to\overline{\nu}_\tau$, $V_{us}\to V_{us}^*$, $\epsilon_j\to\epsilon_j^*$, ($j=S,T$). Thus, one finds
\begin{equation}
<\cos\alpha>_i^{\tau^-}+<\cos\alpha>_i^{\tau^+}=\frac{\int_{s_1,i}^{s_2,i}A(s)\mathrm{d}s}{\int_{s_1,i}^{s_2,i}E(s)\mathrm{d}s},\;\quad <\cos\alpha>_i^{\tau^-}-<\cos\alpha>_i^{\tau^+}=\frac{\int_{s_1,i}^{s_2,i}B(s)\mathrm{d}s}{\int_{s_1,i}^{s_2,i}E(s)\mathrm{d}s}, 
\end{equation}
with
\begin{eqnarray}
&&A(s)\,=\,\frac{G_F^2|V_{us}|^2M_\tau^3S_{EW}}{1024\pi^3s^3}\left(1-\frac{s}{M_\tau^2}\right)^2\lambda(s,m_K^2,m_\pi^2)\Delta_{K\pi}\nonumber\\
&&\times\Bigg\lbrace-\frac{8}{3}\left(1+\frac{\Re e[\hat{\epsilon}_S]s}{M_\tau(m_s-m_u)}\right)\Re e[F_+(s)F_0^*(s)]+\frac{16s}{3M_\tau}\Re e[\hat{\epsilon}_T] \Re e[F_T(s)F_0^*(s)]\Bigg\rbrace\,,\;\;\;\;\;\;\;\;
\end{eqnarray}

\begin{eqnarray}
&&B(s)\,=\,\frac{G_F^2|V_{us}|^2M_\tau^3S_{EW}}{1024\pi^3s^3}\left(1-\frac{s}{M_\tau^2}\right)^2\lambda(s,m_K^2,m_\pi^2)\Delta_{K\pi}\nonumber\\
&&\times\Bigg\lbrace\frac{8}{3}\frac{\Im m[\hat{\epsilon}_S]s}{M_\tau(m_s-m_u)}\Im m[F_+(s)F_0^*(s)]+\frac{16s}{3M_\tau}\Im m[\hat{\epsilon}_T]\Im m[F_T(s)F_0^*(s)]\Bigg\rbrace\,,\;\;\;\;\;\;\;\;
\end{eqnarray}

\begin{eqnarray}
&&E(s)\,=\,\frac{G_F^2|V_{us}|^2M_\tau^3S_{EW}}{1024\pi^3s^3}\left(1-\frac{s}{M_\tau^2}\right)^2\lambda^{1/2}(s,m_K^2,m_\pi^2)\nonumber\\
&&\times\Bigg\lbrace\frac{2}{3}\lambda(s,m_K^2,m_\pi^2)\left(1+\frac{2s}{M_\tau^2}\right)|F_+(s)|^2+2\Delta_{K\pi}^2\Bigg|1+\frac{\hat{\epsilon}_Ss}{M_\tau(m_s-m_u)}\Bigg|^2|F_0(s)|^2\nonumber\\
&&+\frac{8}{3}\lambda(s,m_K^2,m_\pi^2)\left(s|\hat{\epsilon}_T|^2\left(2+\frac{s}{M_\tau^2}\right)|F_T(s)|^2-\frac{3s}{M_\tau}\Re e[\hat{\epsilon}_T F_T(s) F_+^*(s)]\right)\,.\;\;\;\;
\end{eqnarray}
\section{Form-factors input}\label{sec:FFinput}
\hspace{0.5cm}We will use the dispersive form factors that were employed in e. g. ref.~\cite{Gonzalez-Solis:2020jlh}, which bound the modulus of the Wilson coefficients $\hat{\epsilon}_{S,T}^D$ using exclusive semileptonic tau decays into one and two mesons. We will illustrate this hadron input here, taking as an example the simplest di-pion $F_+$ case, \cite{Pich:2001pj,GomezDumm:2013sib,Gonzalez-Solis:2019iod}. The other dispersive form factors in this channel and those for the remaining decay modes were worked out in e. g. refs.~\cite{Descotes-Genon:2014tla, Miranda:2018cpf,Jamin:2000wn,Jamin:2001zq,Moussallam:2007qc,Boito:2008fq,Boito:2010me,Antonelli:2013usa,Bernard:2013jxa,Rendon:2019awg,Escribano:2013bca,Escribano:2014joa,Gonzalez-Solis:2019lze} and agree well with data, given the present uncertainties. We will briefly comment on the sources of theory uncertainty for $F_+^{\pi\pi}$, which is the best known, below.\\

The pion vector form factor is obtained through a three-times subtracted dispersion relation
\begin{equation}
F_+^{\pi\pi}(s)=\mathrm{exp}\left[\alpha_1s+\frac{\alpha_2}{2}s^2+\frac{s^3}{\pi}\int_{4m_\pi^2}^\infty\frac{\mathrm{d}s'}{s'^3}\frac{\delta_+(s')}{s'-s-i0}\right]\,,
\end{equation}
where $\alpha_{1,2}$ are subtraction constants (the other one is fixed by the conservation of the vector current, so that $F_+(0)=1$) giving the slope and curvature of the threshold expansion of $F_+(s)$, where Chiral Perturbation Theory provides useful constraints \cite{Weinberg:1978kz, Gasser:1983yg, Gasser:1984gg, Cata:2007ns}. In practice, since the phaseshift $\delta_+(s)$ is not known up to infinity, several approaches are used to account for the associated uncertainty, as detailed in the quoted refs. In the resonance region we obtain the seed for $\tan\delta_+(s)=\frac{\Im m F_+(s)}{\Re e F_+(s)}$ from Resonance Chiral Theory ($R\chi T$)~\cite{Ecker:1988te,Ecker:1989yg}, with parameters encoding the resonance properties (their pole positions and interferences among them) that are fitted to data. Our form factors comply with the leading asymptotic behaviour predicted by perturbative QCD \cite{Lepage:1980fj}.\\

In the remainder of this section, we summarize the main sources of theory uncertainty, analyzed minutely in ref.~\cite{Gonzalez-Solis:2019iod} for $F_+^{\pi\pi}(s)$. In the elastic zone (which, neglecting the phase-space suppressed four-pion contributions extends up to the di-Kaon channel, with invariant mass around a GeV), the form factor phase is known with extreme precision, thanks to the Roy-Steiner equation analyses of pion scattering data by the Madrid \cite{Garcia-Martin:2011iqs} and Bern \cite{Caprini:2011ky} groups (and to the Watson theorem \cite{Watson:1952ji} that ensures the equality of the scattering and form factor phases in this region). Ref.~\cite{Gonzalez-Solis:2019iod}, working within the excellent approximation of isospin symmetry, uses these results in the elastic domain, including the possible small variation of the precise energy at which the phase reaches $\pi/2$, signaling the $\rho(770)$ resonance position. Another source of uncertainty corresponds to the matching point between the elastic phase and the one obtained using as seed the $R\chi T$ expression, where values in the interval $[0.9,1]$ GeV are employed. For the $\rho^{\prime(\prime)}$ resonances widths, both the simplest case with only the di-pion channel and also that adding the di-Kaon and $\omega$-$\pi$ cuts were considered. As commented before, there is a lack of experimental information on $\delta_+^{\pi\pi}(s)$ above a given energy, and we only know that it must approach $\pi$ asymptotically~\cite{Leutwyler:2002hm}. Several methods quantified this uncertainty in ref.~\cite{Gonzalez-Solis:2019iod}: cutting the upper limit ($s_{cut}$) of the dispersive integral in $4$, or $9$ GeV$^2$, or keeping it up to $\infty$; modifying the phase for $s\sim s_{cut}$ to avoid spurious singularities; including additional inelastic contributions via a conformal polynomial, etc. It was also verified that the number of subtraction constants was optimal using the reduced $\chi^2$ as a criterion. The impact of the $\tau^-\to K^-K_S\nu_\tau$ data was also explored, although it is not yet precise enough to improve our knowledge of the $\rho^\prime-\rho^{\prime\prime}$ interference.\\

\begin{figure}[h!]
\includegraphics[width=10cm]{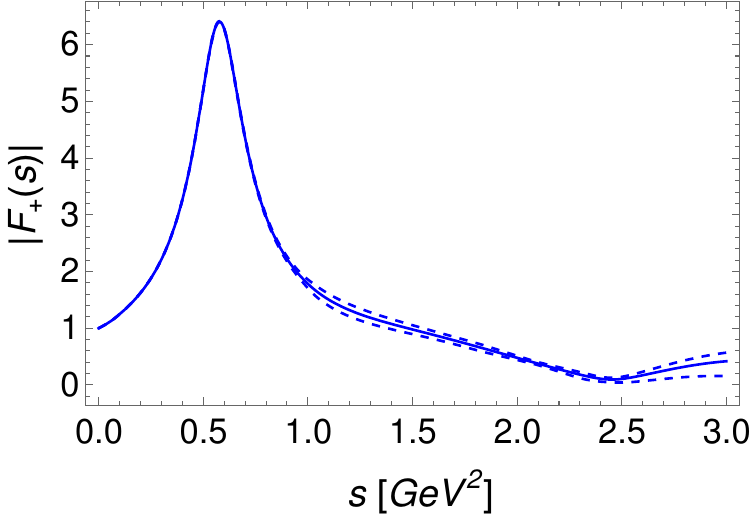}
\centering
\caption{Pion vector form factor modulus as a function of the $\pi\pi$ invariant mass. The solid curve represents the central value, and the dashed lines cover the one sigma uncertainties, according to ref.~\cite{Gonzalez-Solis:2019iod}.}\label{Fig_AbsF+}
\end{figure}

\begin{figure}[h!]
\includegraphics[width=10cm]{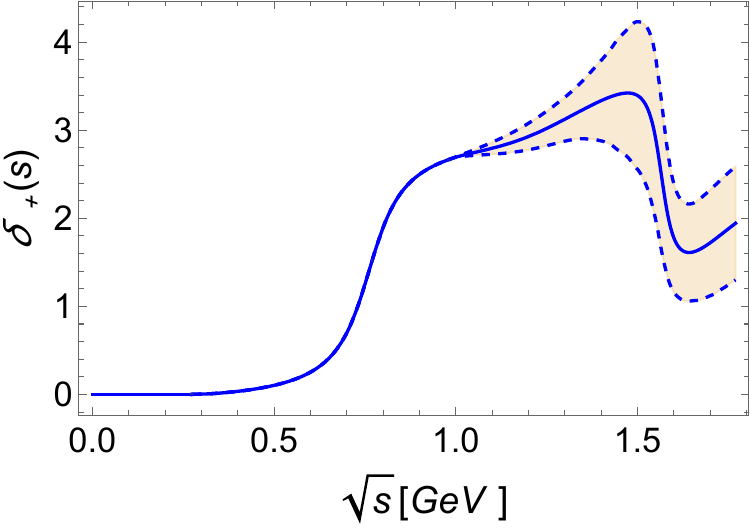}
\centering
\caption{Pion vector form factor phase as a function of the $\pi\pi$ invariant mass. The solid curve represents the central value, and the dashed lines cover the one sigma uncertainties, according to ref.~\cite{Gonzalez-Solis:2019iod}.}\label{Fig_deltaF+}
\end{figure}

\begin{figure}[h!]
\includegraphics[width=10cm]{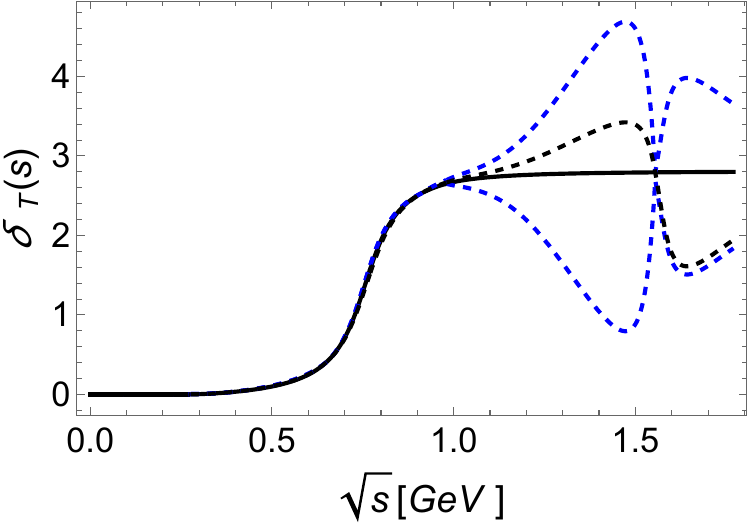}
\centering
\caption{Pion tensor form factor phase as a function of the $\pi\pi$ invariant mass, assuming $|\delta_+(s)-\delta_T(s)|=2\delta_+^{inel}(s)$. The solid curve corresponds to the central value (neglecting inelastic contributions), and the black dashed inner line depicts $\delta_+(s)$ for comparison, according to ref.~\cite{Gonzalez-Solis:2019iod} (see figure~\ref{Fig_deltaF+}).}\label{Fig_deltaFT}
\end{figure}

\begin{figure}[h!]
\includegraphics[width=10cm]{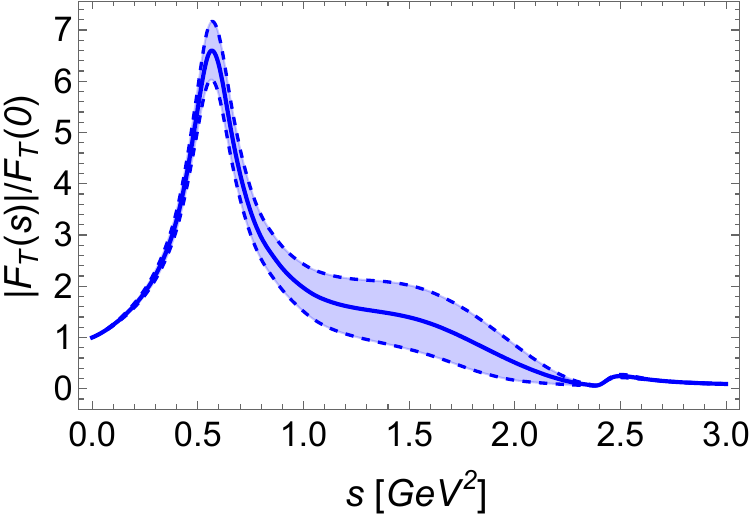}
\centering
\caption{Pion tensor form factor modulus as a function of the $\pi\pi$ invariant mass. The solid curve corresponds to the central value, and the dashed ones depict one standard deviation uncertainties, assuming $|\delta_+(s)-\delta_T(s)|=2\delta_+^{inel}(s)$. }\label{Fig_AbsFT}
\end{figure}

In this way, ref.~\cite{Gonzalez-Solis:2019iod} quantified the uncertainty of $F_+^{\pi\pi}(s)$, whose modulus and phase are represented in figures~\ref{Fig_AbsF+} and \ref{Fig_deltaF+}. Following ref.~\cite{Cirigliano:2017tqn}, we estimated the difference between the vector and tensor form factors in the inelastic region~\footnote{They coincide in the elastic one, as elegantly explained in ref.~\cite{Cirigliano:2017tqn}.} ($inel$) using $\delta_+(s)-\delta_T(s)=2\delta_+^{inel}(s)$. The tensor form factor phase, $\delta_T(s)$, thus obtained is compared in figure~\ref{Fig_deltaFT} to $\delta_+(s)$. Figure~\ref{Fig_AbsFT} plots $|F_T(s)|/F_T(0)$. In this case, we lack information on the slope parameters, so we use a once-subtracted dispersion relation, which increases the sensitivity to $s_{cut}$. As a result, the uncertainty on $|F_T(s)|$ is larger than for $|F_+(s)|$, despite being their phases identical in the elastic regime.\\


\section{Bounds on the New Physics coefficients}\label{sec:BoundsonEpsilons}
\hspace{0.5cm} The Lagrangian eq.~(\ref{LEFTLagrangian}) is obtained as the relevant part of the low-energy limit of the SMEFT Lagrangian, integrating out the heavy particles, that are non-dynamical at the considered scales. This connection is useful to constrain the imaginary part of the scalar and tensor Wilson coefficients, as we explain in the following (see also refs.~\cite{Cirigliano:2017tqn, Chen:2021udz}).

We will start with the tensor operators. In the gauge basis one has
 \begin{eqnarray}
&& \mathcal{L}_{SMEFT} \supset [C^{(3)}_{\ell e q u}]_{klmn}(\bar{\ell}^i_{Lk}\sigma_{\mu\nu}e_{Rl})\epsilon^{ij}(\bar{q}^j_{Lm}\sigma^{\mu\nu}u_{Rn})+\mathrm{h.c.}\nonumber\\
&& = [C^{(3)}_{\ell e q u}]_{klmn}[(\bar{\nu}_{Lk}\sigma_{\mu\nu}e_{Rl})(\bar{d}_{Lm}\sigma^{\mu\nu}u_{Rn})-(\bar{e}_{Lk}\sigma_{\mu\nu}e_{Rl})(\bar{u}_{Lm}\sigma^{\mu\nu}u_{Rn})]+\mathrm{h.c.},\;\;\;\;\;\;\;\;\;\;
 \end{eqnarray}
 where the left-handed lepton and quark $SU(2)_L$ doublets are $\ell_L=(\nu_L,e_L)^T$ and $q_L=(u_L,d_L)^T$ and the corresponding singlets are $e_R$ and $u_R$. $SU(2)_L$(family) indices are $i,j(k,l,m,n)$, respectively. In the \textit{down} mass basis,
 \begin{eqnarray}\label{eq_SMEFT_MB}
 \mathcal{L}_{SMEFT}&&\supset [C^{(3)}_{\ell e q u}]_{klmn}[(\bar{\nu}_{Lk}\sigma_{\mu\nu}e_{Rl})(\bar{d}_{Lm}\sigma^{\mu\nu}u_{Rn})-V_{am}(\bar{e}_{Lk}\sigma_{\mu\nu}e_{Rl})(\bar{u}_{La}\sigma^{\mu\nu}u_{Rn})]\nonumber \\&&\quad+\mathrm{h.c.}
 \end{eqnarray}
 In the previous Lagrangian, eq.~(\ref{eq_SMEFT_MB}), we will be interested in the case $k=l=3$ ($\tau$ flavor). In the quark sector, $n=1$ and $m=1,2$ will be relevant for the first term (depending on $D$ being $d,s$). In the second one also $n=2$ will be important in the constraint coming from $D^0-\bar{D}^0$ mixing, so $a,m=1,2$ will matter.
 The LEFT and SMEFT coefficients are related at leading order by
 \begin{equation}
 [C^{3}_{\ell equ}]_{33mn}=-2\sqrt{2}G_FV_{uD}\hat{\epsilon}^{D*}_T\,.
 \end{equation}
\\
\indent The operators in the second term of eq.~(\ref{eq_SMEFT_MB}) contribute to the neutron electric dipole moment (through the $u$-quark contribution to it) and to $D^0-\bar{D}^0$ mixing, which constrain stringently $\Im m[\epsilon_T^D]$. We will consider, in our numerical analysis, two scenarios, as in ref.~\cite{Chen:2021udz}:
 \begin{itemize}
\item If the neutron EDM~\footnote{We recall that \cite{Chen:2021udz} $d_n=g_T^u(\mu)d_u(\mu)$, with $g_T^u(2\mathrm{\,GeV})=-0.204(14)$ \cite{Bhattacharya:2015esa, Gupta:2018lvp}, $|d_n|<1.8\times10^{-26}e$ cm at $90\%$ confidence level, C. L. \cite{Abel:2020pzs}.} is mainly contributed by a single $\hat{\epsilon}_T^D$, then
\begin{equation}
d_u(\mu)=-2\sqrt{2}G_F\frac{eM_\tau}{\pi^2}V_{uD}^2\Im m[\hat{\epsilon}^D_T(\mu)]\log\frac{\Lambda}{\mu}\,,
\end{equation}
which yields $|\Im m[\hat{\epsilon}_T^s]|\lesssim4\times10^{-6}$ and $|\Im m[\hat{\epsilon}_T^d]|\lesssim8\times10^{-5}$ for $\Lambda\gtrsim 100$ GeV and $\mu=2$ GeV~\footnote{The relative strength of these bounds is explained by the small $|V_{us}/V_{ud}|^2$ ratio.}. The constraint on $D^0-\bar{D}^0$ mixing will then come from $\phi=\Im m\Big [V_{uD}V_{cD}\epsilon_{T(S)}^D\Big]$.
\item If both terms contribute, then the corresponding restriction reads
\begin{equation}
d_u(\mu)=-2\sqrt{2}G_F\frac{eM_\tau}{\pi^2}(V_{ud}^2 \Im m[\hat{\epsilon}_T^d(\mu)]+V_{us}^2 \Im m[\hat{\epsilon}_T^s(\mu)])\log\frac{\Lambda}{\mu}\,.
\end{equation}
In this case the constraint from $D^0-\bar{D}^0$ mixing will apply to $\phi=\Im m\Big [V_{ud}V_{cd}\epsilon_{T(S)}^d+V_{us}V_{cs}\epsilon_{T(S)}^s\Big]$
 \end{itemize}
 We will consider $\phi=\pm\frac{\pi}{4}$, as in refs.~\cite{Cirigliano:2017tqn, Chen:2021udz}.\\

 We turn now to the scalar operators in the SMEFT. We have
 \begin{equation}
\mathcal{L}_{SMEFT}\supset [C^{(1)}_{\ell e q u}]_{klmn}\left(\bar{\ell}^i_{Lk}e_{Rl}\right)\epsilon^{ij}\left(\bar{q}^j_{Lm}u_{Rn}\right)+[C_{\ell e d q}]_{klmn}\left(\bar{\ell}^i_{Lk}e_{Rl}\right)\left(\bar{d}_{Rm}q^i_{Ln}\right)+\mathrm{h.c.}\,,
 \end{equation}
 which, in the fermion mass basis, reads
  \begin{eqnarray}
\mathcal{L}_{SMEFT}&\supset& [C^{(1)}_{\ell e q u}]_{klmn}\left[\left(\bar{\nu}_{Lk}e_{Rl}\right)\left(\bar{d}_{Lm}u_{Rn}\right)-V_{am}(\bar{e}_{Lk}e_{Rl})(\bar{u}_{La}u_{Rn})\right]\nonumber\\
&+&[C_{\ell e d q}]_{klmn}\left[V_{an}^*\left(\bar{\nu}_{Lk}e_{Rl}\right)\left(\bar{d}_{Rm}u_{La}\right)+\left(\bar{e}_{Lk}e_{Rl}\right)\left(\bar{d}_{Rm}d_{Ln}\right)\right]+\mathrm{h.c.}\;\;\;\;\;\;\;\;\;\;\;
 \end{eqnarray}
 Here we will again be interested in the case with $k=l=3$, and $m,n,a=1,2$. LEFT and SMEFT coefficients are related by (we neglect the contribution proportional to $V_{ub}^*$)
 \begin{equation}
-2\sqrt{2}G_FV_{uD}^* \hat{\epsilon}_S^{D*}=[C^{(1)}_{\ell e q u}]_{33m1}+V_{ud}^*[C_{\ell e d q}]_{33m1}+V_{us}^*[C_{\ell e d q}]_{33m2}
.
 \end{equation}
 Although the scalar operators do not enter directly the neutron EDM, they are still restricted indirectly because the bound on the tensor coefficients impacts the scalar ones through renormalization group equation evolution (details are given in ref.~\cite{Chen:2021udz}). If we keep for simplicity only $C^{(1)}_{\ell e q u}$ \cite{Chen:2021udz}, $|\Im m[\hat{\epsilon}_T^s]|\lesssim4\times10^{-6}\Rightarrow|\Im m[\hat{\epsilon}_S^s]|\lesssim2\times10^{-3}$ and $|\Im m[\hat{\epsilon}_T^d]|\lesssim8\times10^{-5}\Rightarrow |\Im m[\hat{\epsilon}_S^d]|\lesssim4\times10^{-2}$.
 In the scalar case the strongest constraints come from $D^0-\bar{D}^0$ mixing, implying \cite{Chen:2021udz} $\Im m\left[\hat{\epsilon}^D_S\right]\in[-3.1,1.6]\times10^{-4}$ at $95\%$ C.L., that is the restriction that dominates.\\

 The real parts of the Wilson coefficients $\epsilon_{T(S)}^s $ are most effectively constrained by analyzing CP-conserving inclusive and exclusive semi-leptonic tau decays~\cite{Chen:2021udz,Garces:2017jpz,Cirigliano:2017tqn,Miranda:2018cpf,Rendon:2019awg,Gonzalez-Solis:2019lze,Gonzalez-Solis:2020jlh,Cirigliano:2018dyk,Cirigliano:2021yto,Arteaga:2022xxy,Escribano:2023seb}. For the strangeness-changing sector, we will take the results from ref.~\cite{Escribano:2023seb} that include improved radiative corrections (see also refs.~\cite{Arroyo-Urena:2021nil,Arroyo-Urena:2021dfe}), which are
 \begin{equation}
 \Re e[\epsilon_S^s]=\left(0.8\pm0.9\right)\times10^{-2}\,,\quad \Re e[\epsilon_T^s]=\left(0.5\pm0.8\right)\times10^{-2}\,,
 \end{equation}
 with a negligible correlation coefficient of $-0.057$.
 
 The real parts of the Wilson coefficients $\epsilon_{T(S)}^{d} $, entering strangeness-conserving processes, were better constrained in \cite{Cirigliano:2018dyk} 
 \begin{equation}
 \Re e[\epsilon_S^d]=\left(-0.6 \pm 1.5\right)\times10^{-2}\,,\quad \Re e[\epsilon_T^d]=\left(-0.04\pm0.46\right)\times10^{-2}\,,
 \end{equation}
 with a  null correlation coefficient.

\section{Phenomenological analysis}\label{sec_Pheno}
\hspace{0.5cm} Here we will present the numerical analysis of the most interesting observables sensitive to  CP violation recalled in section \ref{sec:LEFTCPVObservables}, using the form factors introduced in section \ref{sec:FFinput}, and respecting the bounds on the Wilson coefficients of section \ref{sec:BoundsonEpsilons}. We will not discuss the $K_S\pi$ case, which has been studied extensively in e.g. refs.~\cite{Chen:2021udz,Cirigliano:2017tqn,Rendon:2019awg}.~\footnote{We have nevertheless reproduced figure 3 in ref.~\cite{Chen:2021udz}, as this observable was not considered in ref.~\cite{Rendon:2019awg}. We thank Xin-Qian Li for useful feedback concerning this verification.} We will present the rate CP and the forward-backward asymmetries for the $\pi^\pm\pi^0$, $K^\pm K_S$ and $K^\pm\pi^0$ modes, in turn. Our results for $A_{CP}^{rate}$ will be quoted for the maximum possible signal, corresponding to $\Im m[\hat{\epsilon}_T^{d}] =\pm8\times10^{-5}$ and $\Im m[\hat{\epsilon}_T^{s}] =\pm4\times10^{-6}$ 
 at $90\%$ C.L. In the case of $A_{FB}$ we will determine the optimal values of the $\hat{\epsilon}_{S,T}^D$ demanding that they maximize the following figure of merit~\footnote{According to the preceding discussion we would naively expect that, in addition to the limits on $\Im m[\hat{\epsilon}_S^{D}]$, given at 90\%C.L. by the interval $[-2.7,1.2]\times10^{-4} $ --which determine $A_{CP}^{rate}$-- the maximal CP violating signals would be found with the following Wilson coefficients  (obtained from the extremes of the $90\%$C.L. intervals) $\Re e[\hat{\epsilon}_S^{d}]= -3.1\times10^{-2}$, $\Re e[\hat{\epsilon}_S^{s}]= 2.3\times10^{-2}$ , for the scalar non-standard interactions, and $\Re e[\hat{\epsilon}_T^{d}]= -7.9\times10^{-3}$ and $\Re e[\hat{\epsilon}_T^{s}]= 1.8\times10^{-2}$, for the tensor interactions.}
 \begin{equation}\label{eq_FOM}
    F(\hat{\epsilon}_{S}^{D},\hat{\epsilon}_{T}^{D}) = \int_{(m_{1} + m_{2})^2}^{M_{\tau}^2}  \Big|A_
    {FB}(s;\hat{\epsilon}_{S}^{D},\hat{\epsilon}_{T}^{D})\Big| \quad  ds\,.
\end{equation}\\
 \subsection{CP violating asymmetries in the $\pi^\pm\pi^0$ channel}\label{sec_pipi}
\hspace{0.5cm} In this case the SM contribution is negligible, so using eq.~(\ref{eq_ACPrateNP}) for the NP part, we find that
\begin{equation} \label{ACPratepipi}
 A_{CP}^{rate}|_{\pi\pi}\leq3\times10^{-5}\,,
 \end{equation}
which is below the expected sensitivity of current and near-future experiments.\\
 
\begin{figure}[h!] 
\includegraphics[width=0.9\linewidth, height=6cm]{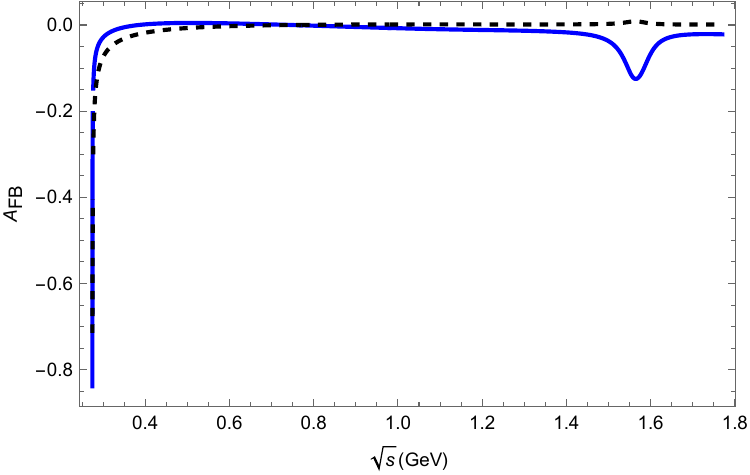} 

\caption{Maximal $A^{\tau\rightarrow \pi\pi^{0}\nu_{\tau}}_{FB}(s) $ (blue solid line), corresponding to the Wilson coefficients values $ \Re e[\epsilon_S^d]=-3.1\times10^{-2}  $ , $ \Im m[\epsilon_S^d]=-2.7\times 10^{-4} $,   $ \Re e[\epsilon_T^d]=-7.9\times 10^{-3}  $ and  $ \Im m[\epsilon_T^d]=8 \times 10^{-5}$, compared to the SM case (black dashed line).}
\label{fig_AFBPiPi}
\end{figure}

 The maximal $A_{FB}|_{\pi\pi}$ is shown by a blue solid line in figure \ref{fig_AFBPiPi}, which was obtained with the values $ \Re e[\epsilon_S^d]=-3.1\times10^{-2} $ , $ \Im m[\epsilon_S^d]=-2.7\times 10^{-4} $,   $ \Re e[\epsilon_T^d]=-7.9\times 10^{-3}  $ and  $ \Im m[\epsilon_T^d]=8 \times 10^{-5}$. For comparison, the SM contribution (very close to zero, except at threshold) is shown by the black dashed line. Since $|A_{FB}(s)|\leq 0.1$ for most of the spectrum, it will be challenging to measure such asymmetries at Belle-II and future experiments.\\

\begin{figure}[h!]

\includegraphics[width=1.2\linewidth, height=8cm]{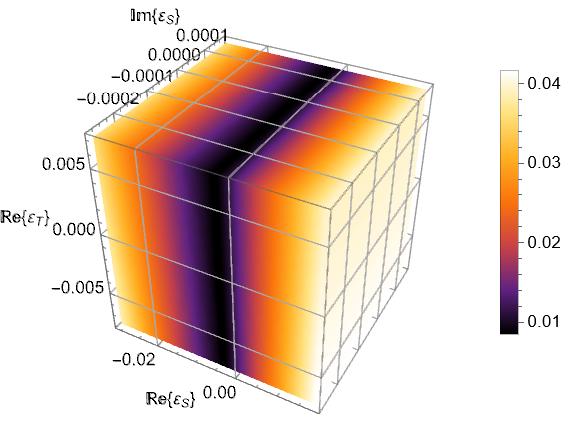}
 
\caption{3D density plot for the FoM described in eq. (\ref{eq_FOM}) on the parameter space, setting $|\Im m[\hat{\epsilon}_T^d]| = 8\times 10^{-5} $, for the $\pi\pi^0$ decay channel. }\label{3Dpipi}
\end{figure}

In the SM, the FoM $F(\hat{\epsilon}_S=0,\hat{\epsilon}_T=0)$ reads $1.2\times10^{-2}$. Allowing the Wilson coefficients to vary within their allowed intervals, we find that its maximum value is $4.1\times10^{-2}$ , which is obtained for $\Re e[\epsilon_S^d]=-3.1\times10^{-2}  $, $ \Im m[\epsilon_S^d]=1.7\times 10^{-4} $, $ \Re e[\epsilon_T^d]=-7.9\times 10^{-3}  $ and  $ \Im m[\epsilon_T^d]=-8 \times 10^{-5}$. Even for this maximal NP allowed, measurement will be challenging. Figure~\ref{3Dpipi} shows a three-dimensional density plot of the FoM $F$ for fixed $ \Im m[\hat{\epsilon}_T^d]$, varying its real part and the complex number $\hat{\epsilon}_S^d$.

 \subsection{CP violating asymmetries in the $K^\pm K_S$ channel}\label{sec_KK}
\hspace{0.5cm} As we will see, this is the most interesting channel. Using eq.~(\ref{eq_FullACPrate}) and the negative bound for $\Im m[\epsilon_T^d] $, allows us to get the maximum absolute value for $A_{CP}^{rate}|_{KK} $, in order to obtain \footnote{We note that, for the $K_SK^\pm$ channel, the expected standard model contribution $A_{CP}^{rate}|_{KK,SM} $ is opposite in sign to that of the channel $K_{S}\pi^{\pm} $.}
 \begin{equation} \label{ACPrateKK}
 A_{CP}^{rate}|_{KK}=-3.83\times10^{-3},
 \end{equation}
 that is dominated by the SM contribution (for which we used the result adapted to the BaBar setting, with a SM CP rate asymmetry of $-3.6\times10^{-3}$) and
 \begin{equation} \label{ACPrateKKNP}
 |A_{CP}^{rate}|_{KK,NP}|\leq 2.3\times10^{-4},
 \end{equation}
 found with the analog for this channel of eq.~(\ref{eq_ACPrateNP}).
 Actually, by considering the interval in (\ref{ACPrateKKNP}), we can set lower and upper bounds for $A_{CP}^{rate}|_{KK} $, \textit{i.e.}
 
 \begin{equation}
     -3.83\times10^{-3}\leq A_{CP}^{rate}|_{KK} \leq -3.37\times 10^{-3}.
 \end{equation} 
Noticeably, a $5\%$ precision on the measurement of $A_{CP}^{rate}$ would already be sensitive to the maximum allowed NP contribution in this case. This contrasts sharply with the $K_S \pi^\pm$ channels, where the NP contributions is orders of magnitude smaller than the SM one, complicating the explanation of the BaBar anomaly. There are two reasons why this NP contribution can be so much larger than for the $K_S \pi^\pm$ modes. On the one hand, the bound on the Wilson coefficient is a factor twenty larger here ($|\Im m[\hat{\epsilon}_T^d]|\leq 8\times10^{-5}$ vs. $|\Im m[\hat{\epsilon}_T^s]|\leq 4\times10^{-6}$).~\footnote{We recall that $A_{CP}^{rate}$ is linear on $\Im m[\hat{\epsilon}_T^D]$, eq.~(\ref{eq_ACPrateNP}).} On the other, the $K_S K^\pm$ channel lies in the inelastic region, while the main contribution to the decay width of the $K_S \pi^\pm$ belongs to the elastic regime, where $A_{CP}^{rate}$ vanishes identically. Due to these facts, we get such a large NP effect, that can be up to $\mathcal{O}(5\%)$ on $A_{CP}^{rate}|_{KK}$. We therefore encourage the experimental collaborations to measure this observable trying to reach this level of precision. On the other hand, a measurement with $10\%$ accuracy of $A_{CP}^{rate}$ in the $K_S K$ channels should coincide (up to sign) with the result for $K_S \pi$, so it will be a purely experimental test of the BaBar anomaly that we prompt BaBar and Belle(-II) to perform. As a reference, in the BaBar measurement \cite{BaBar:2018qry}, the uncertainty of the best measured points is $\sim 3\%$, so these numbers are not unrealistic targets.\\
 
\begin{figure}[h!] 
\includegraphics[width=10cm]{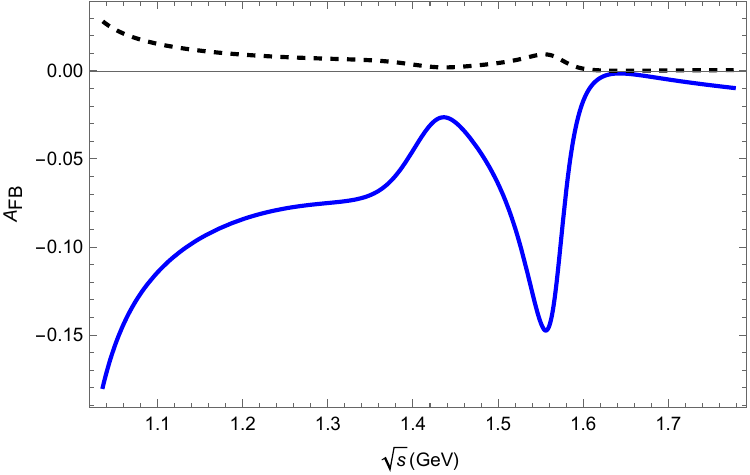}
\centering
\caption{Maximal $A^{\tau\rightarrow K_SK\nu_{\tau}}_{FB}(s) $ (blue solid line), corresponding to the Wilson coefficients values $\Re e[\epsilon_S^d]=-3.1\times10^{-2} $ , $ \Im m[\epsilon_S^d]=-2.7\times 10^{-4} $,   $ \Re e[\epsilon_T^d]=7.1\times 10^{-3}  $ and  $ \Im m[\epsilon_T^d]=8 \times 10^{-5}$, compared to the SM case (black dashed line).}\label{fig_AFBKK}
\end{figure}

 The maximal $A_{FB}|_{KK}$ is shown by a blue solid line in figure \ref{fig_AFBKK}, which is obtained with the set of Wilson coefficients, $\Re e[\epsilon_S^d]=-3.1\times10^{-2}  $ , $ \Im m[\epsilon_S^d]=-2.7\times 10^{-4} $,   $ \Re e[\epsilon_T^d]=-7.1\times 10^{-3}  $ and  $ \Im m[\epsilon_T^d]=8 \times 10^{-5}$. The (unmeasurably small) SM contribution is represented by a black dashed line. The obtained asymmetries, of the order of $-0.1$ in a big portion of the spectrum, should be observable at current and future facilities.\\

\begin{figure}[h!]

\includegraphics[width=1.2\linewidth, height=8cm]{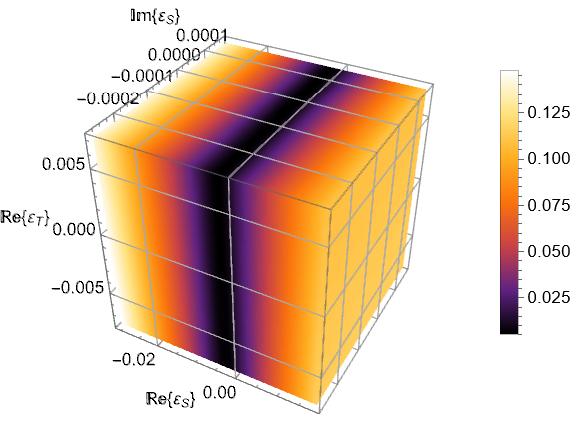}
 
\caption{3D density plot for the FOM described in eq. (\ref{eq_FOM}) on the parameter space, fixing $|\Im m[\hat{\epsilon}_T^d]| = 8\times 10^{-5} $,  for the $KK_S$ decay channel. }\label{3DKK}
\end{figure}

In the SM, the FoM $F(\hat{\epsilon}_S=0,\hat{\epsilon}_T=0)$ reads $1.1\times10^{-2}$. Varying the Wilson coefficients within their allowed ranges, we find that its maximum value is $0.14$, which is obtained for $\Re e[\epsilon_S^d]=-3.1\times10^{-2} $, $ \Im m[\epsilon_S^d]=-2.7\times 10^{-4} $, $ \Re e[\epsilon_T^d]=7.1\times 10^{-3}  $ and  $ \Im m[\epsilon_T^d]=8 \times 10^{-5}$. For this maximal NP allowed, measurement shall be possible at present experiments. Figure~\ref{3DKK} shows a three-dimensional density plot of the FoM $F$ for fixed $ \Im m[\hat{\epsilon}_T^d]$, varying its real part and the complex number $\hat{\epsilon}_S^d$.\\

 \subsection{CP violating asymmetries in the $K^\pm\pi^0$ channel}\label{sec_Kpi}
\hspace{0.5cm} From eq.~(\ref{eq_ACPrateNP}) we compute
 \begin{equation} \label{ACPrateKpi}
 A_{CP}^{rate}|_{K\pi}\leq6\times10^{-7}\,,
 \end{equation}
 that is again too small to be probed soon.\\
 
\begin{figure}[h!]
\includegraphics[width=10cm]{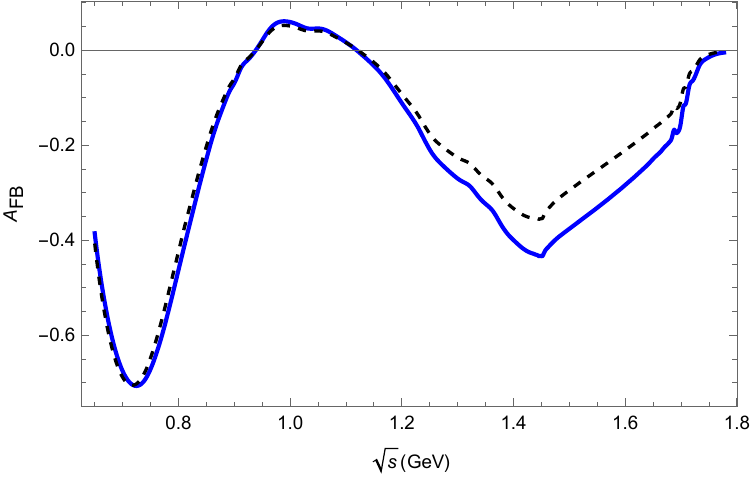}
\centering
\caption{Maximal $A^{\tau\rightarrow K\pi^0\nu_{\tau}}_{FB}(s) $ (blue solid line), corresponding to the Wilson coefficients values $ \Re e[\epsilon_S^s]=2.3\times 10^{-2}  $ , $ \Im m[\epsilon_S^s]=-2.7\times 10^{-4} $,   $ \Re e[\epsilon_T^s]=1.8    \times 10^{-2}  $ and  $ \Im m[\epsilon_T^s]=4 \times 10^{-6}  $, compared to the SM case (black dashed line).}\label{fig_AFBKpi}
\end{figure}

 The maximal $A_{FB}|_{K\pi}$ is shown by a solid blue line in figure \ref{fig_AFBKpi}, which was obtained for the following values of the Wilson coefficients:  $ \Re e[\epsilon_S^s]=2.3\times 10^{-2}  $ , $ \Im m[\epsilon_S^s]=-2.7\times 10^{-4} $,   $ \Re e[\epsilon_T^s]=1.8    \times 10^{-2}  $ and  $ \Im m[\epsilon_T^s]=4 \times 10^{-6}  $. The SM contribution is depicted by a black dashed line. The small difference between them seems to prevent any observation of NP in this observable in current or forthcoming experiments. In this case both curves are clearly non-vanishing, for most of the phase space, reflecting the effect of the kaon-pion mass difference on the angular distribution (as opposed to the channels discussed in sections \ref{sec_pipi} and \ref{sec_KK}, where this difference vanishes in the very good approximation of isospin symmetry).\\

\begin{figure}[h!]
\includegraphics[width=0.9\linewidth, height=6cm]{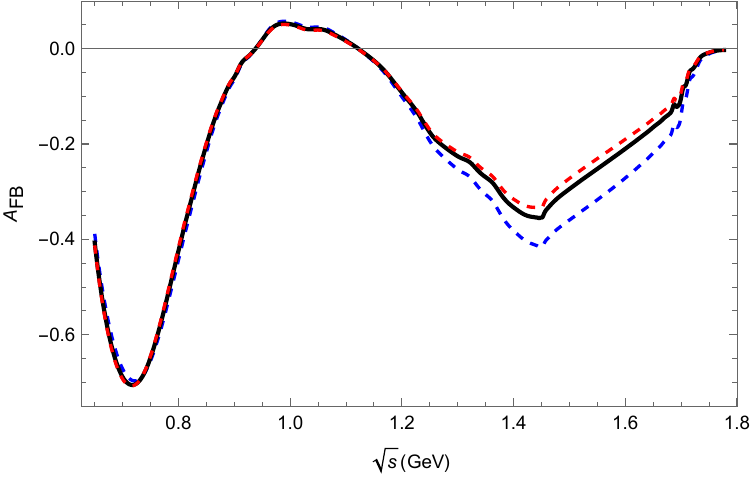} 

\caption{$A^{\tau\rightarrow K\pi^{0}\nu_{\tau}}_{FB}(s) $ in the SM (black solid curve) is compared to the set of allowed Wilson coefficients values that maximize and minimize the FoM $F$, shown as dashed lines.}\label{Fig_3curves}
\end{figure}

\begin{figure}[h!]

\includegraphics[width=1.2\linewidth, height=8cm]{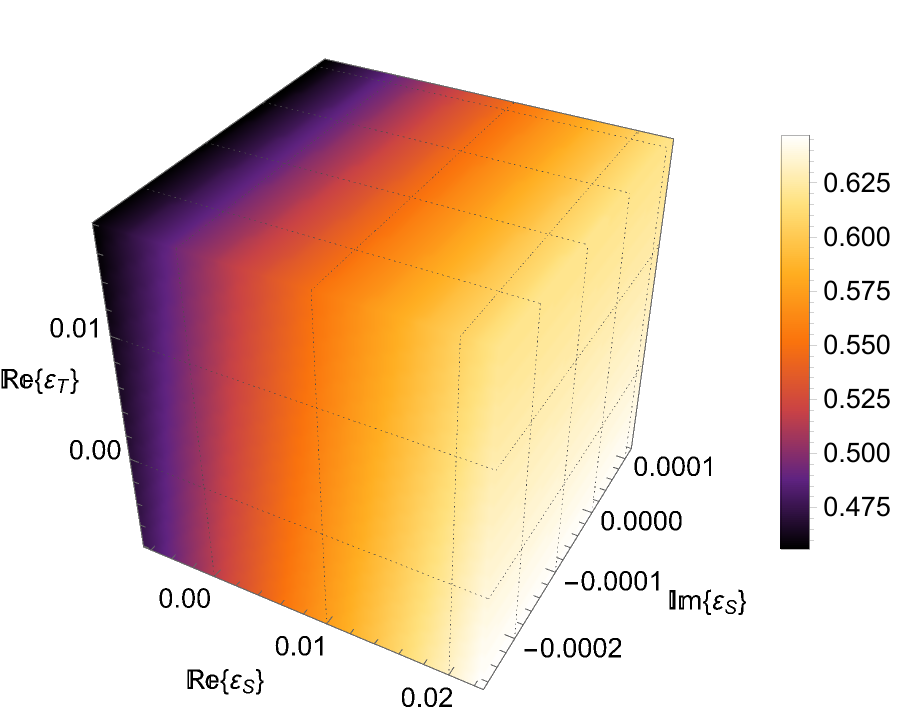}
 
\caption{3D density plot for the FoM described in eq. (\ref{eq_FOM}) on the parameter space, with $|\Im m[\hat{\epsilon}_T^d]| = 4\times 10^{-6} $, for the decay channel with $K^{-}  $ and $\pi^{0} $ in the final state. }\label{3DKpi}
\end{figure}

In the SM, the FoM $F(\hat{\epsilon}_S=0,\hat{\epsilon}_T=0)$ reads $0.45$. Varying the Wilson coefficients within their allowed ranges, we find that its maximum value is $0.64$, which is obtained for $\Re e[\epsilon_S^s]=2.3\times10^{-2}  $, $ \Im m[\epsilon_S^d]=-2.7\times 10^{-4} $, $ \Re e[\epsilon_T^d]=-8\times 10^{-3}  $ and  $ \Im m[\epsilon_T^d]=-4 \times 10^{-6}$. For this maximal CP violation allowed, measurement shall be possible at present experiments. Although it may seem that differentiation from the SM result is fairly achievable, it would actually be challenging, according to Figure~\ref{Fig_3curves}, since it would require a quite accurate measurement in the $K^*(1410)$ region. Figure~\ref{3DKpi} shows a three-dimensional density plot of the FoM $F$ for fixed $ \Im m[\hat{\epsilon}_T^s]$, varying its real part and the complex number $\hat{\epsilon}_S^s$.\\

For the $K\eta^{(\prime)}$ channels we obtain small $A_{CP}^{rate}$ ($\lesssim 5\times10^{-6}$) and non-negligible deviations due to NP in $A_{FB}$ and the FoM $F$~\footnote{These results will be reported elsewhere~\cite{DanielThesis}.}. However, since these are inelastic channels for $K\pi^0$, it would be very demanding to control the SM prediction at the required level to claim NP using the asymmetries for these decay modes.

\section{Conclusions}\label{sec_Concl}
\hspace{0.5cm} CP violation remains to be one of the most active areas of study in high-energy physics, mainly due to its cosmological implications regarding, in particular, the birth of the matter-antimatter asymmetry in the Universe.\\

All precision measurements in the quark sector do not seem to require a NP CP-violating source. Although there are compelling hints for CP non-conservation in the leptons, it is still too early to figure out if non-standard sources (possibly related to the neutrino Majorana nature) are at work.\\

Within this scenario, solving the puzzle related to the anomalous BaBar measurement of $A_{CP}^{rate}$ for the $K_S\pi$ channels, at the level of $-3.6\times10^{-3}$ (with opposite sign and equal magnitude to its SM prediction, constituting a $2.8\,\sigma$ discrepancy), can shed light on novel CP violation. Because of this importance, there were a number of devoted studies examining this conundrum using EFTs, concluding that --unless an unnatural fine-tunning takes place-- it is impossible to explain the BaBar measurement as due to NP above the electroweak scale.\\

In this work we have examined --within the same EFT setting-- the rate and forward-backward asymmetries for the other two-meson tau decay modes, focusing on the $(\pi/K)\pi^0$ and $KK_S$ cases. The most promising one turns out to be the di-Kaon channel, where measurements of $A_{CP}^{rate}$ with a $5\%$ precision would already be sensitive to NP contributions that are currently allowed. For these modes, also measurements of $A_{FB}$ and its integral in energy (that we quantify by our FoM $F$ introduced in eq.~(\ref{eq_FOM})) would be able to distinguish NP permitted by present data. An additional and very timely interest of measuring $A_{CP}^{rate}$ for the $KK_S$ modes would be checking the result for $\pi K_S$, as they coincide (up to sign) unless ad hoc NP is invoked.\\

In the $\pi\pi^0$ channels $A_{FB}$ and $F$ would again be sensitive to NP in current and planned facilities, although difficult to measure. Disentangling NP from SM contributions would be however (much) more challenging in the $K\pi^0$ ($K\eta^{(\prime)}$) modes. In the first case because these deviations are mostly placed in the $K^*(1410)$ region and towards the end of the spectrum, where there are fewer events and data is less precise; and in the second because of the uncertainties on the $K\eta$ data ($\tau\to K\eta^\prime\nu_\tau$ decays have not been measured yet) and the NP effects being in the inelastic region of the coupled-channels $K(\pi^0/\eta/\eta^\prime)$ problem, which is not so controlled.\\

We hope that our analysis motivates the Tau physics group at Belle-II to tackle these interesting analyses that we are proposing. They will also be attractive for the planned super-tau-charm factories. If future facilities produce polarized taus, this will open a bunch of new CP violating measurements exploiting this feature, which would deserve dedicated theory and sensitivity studies for the corresponding collaboration analyses. In these, again, the use of EFTs would allow to set model-independent constraints on the possible deviations from SM results owing to heavy new particles, that would be most useful for the experimental collaborations.

\section*{Acknowledgements}
\hspace{0.5cm} D. A. L. A. thanks CONAHCYT for the financial support during his Ph.D studies. J. R. acknowledges support from the program 'Estancias posdoctorales por México' of CONAHCYT and also from UNAM-PAPIIT grant number IG100322 and by Consejo Nacional de Humanidades, Ciencia y Tecnología grant number CF-2023-G-433. P.~R. acknowledges Conahcyt (México) funding through project CBF2023-2024-3226 as well as Spanish support during his sabbatical through projects MCIN/AEI/10.13039/501100011033, grant PID2020-114473GB-I00, and Generalitat Valenciana grant PROMETEO/2021/071.

\end{document}